\documentclass[epj]{svjour}

\usepackage{graphicx}
\usepackage{fancyhdr}
\usepackage{amsmath,amssymb}
\def\makeheadbox{{
 }}
\def\mytitle{My title} 
\def\myauthors{My name}  
\def\mytype{My type of session}
\def\mysession{My session}

\def\mytitle{SUSY Multi-Step Unification without Doublet-Triplet
  Splitting}
\def\myauthors{J\"urgen Reuter}    
\def\mytype{Parallel Talk}    
\def\mysession{Theoretical Models}

\begin{document}
\title{SUSY Multi-Step Unification without Doublet-Triplet
  Splitting}
\author{J\"urgen Reuter
\thanks{\emph{Email:} juergen.reuter@desy.de}%
}
\institute{University of Freiburg, Institute
 of Physics, Germany
}
%


\date{}
\abstract{
Matter-Higgs unification in string-inspired supersymmetric Grand
Unified Theories predicts the existence of colored states in the Higgs
multiplets and calls for two extra generations of Higgs-like fields
('unhiggses'). If these states are present near the TeV scale,
gauge-coupling unification points to the existence of two distinct
scales, $10^{15}$ GeV where right-handed neutrinos and a Pati-Salam
symmetry appear, and $10^{18}$ GeV where complete unification is
achieved. Baryon-number conservation, while not guaranteed, can
naturally emerge from an underlying flavor symmetry. Collider
signatures and dark-matter physics may be drastically different from
the conventional MSSM. 
} 
\maketitle
%

\newcommand{\rep}[1]{\mbox{\boldmath$#1$}}%
\newcommand{\arep}[1]{\mbox{\boldmath$\overline{#1}$}}%
\newcommand{\vev}[1]{{\langle #1 \rangle}}

\section{Introduction}

One of the most intriguing ideas for physics beyond the Standard Model
are Grand unified theories (GUT), putting together part of or all SM
particles in multiplets of one unified gauge group that is broken down
to $SU(3)_c \times SU(2)_L \times U(1)_Y$ at the GUT scale
$M_\text{GUT}\approx 10^{16}$ GeV. The oldest approaches were based on
the groups $SU(5)$~\cite{Georgi:1974sy} and $SU(4)\times SU(2)\times
SU(2)$~\cite{PS}. GUT models became much more consistent with the
inclusion of low-scale supersymmetry~\cite{SUSY-GUTs} which stabilizes
the hierarchies of the GUT and electroweak (EW) scale and allows for
an almost exact unification of the gauge couplings. While in the
$SU(5)$ GUTs only a partial unification is achieved, in the Pati-Salam
(PS) symmetry lepton number is treated as a fourth
color. $SO(10)$~\cite{SO10} includes all matter particles of the SM
(including a right-handed neutrino), but it does not unify matter and
Higgs representations. This is achieved within the framework of $E_6$
GUTs~\cite{Gursey:1975ki}.  

All mentioned models share the famous doublet-triplet splitting
problem~\cite{DTS}: embedding all states (including Higgs) in complete
representations implies the existence of a pair of 'exotic'
color-triplet EW-singlet superfields $D$ and $D^c$.
Higgs-matter unification furthermore introduces two extra
Higgs/$D$/$D^c$ generations.  If these have EW-scale masses,
their effect on the running couplings spoils unification.
$D/D^c$ superpotential interactions invariant under the GUT symmetries
contain both diquark and leptoquark couplings and thus induce rapid proton
decay~\cite{protondec}.  To avoid this problem, they are usually
placed near the GUT scale, although this cannot be explained without
further structure beyond the gauge symmetry.

\begin{figure}
\begin{center}
\includegraphics[width=.45\textwidth]{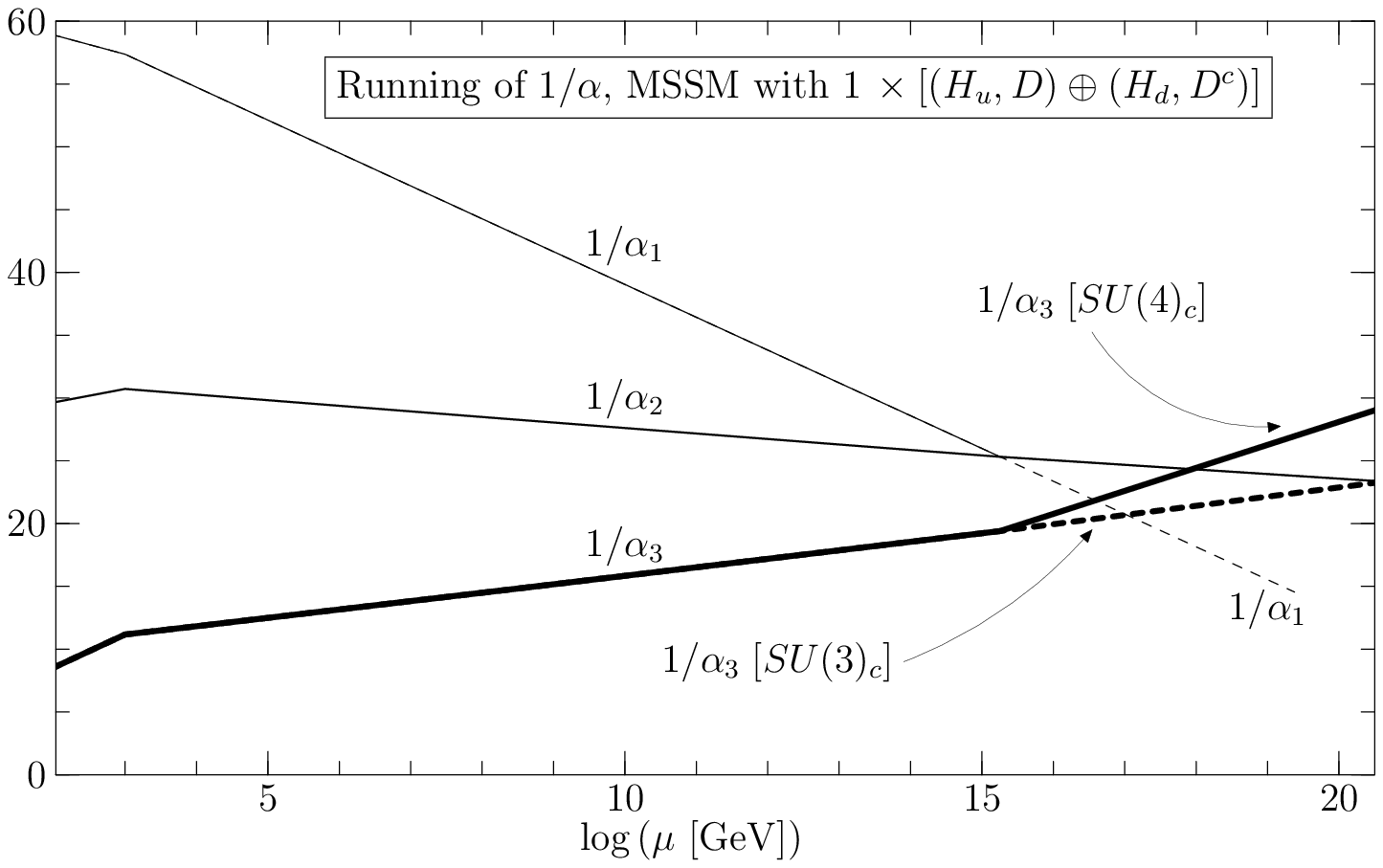}
\includegraphics[width=.45\textwidth]{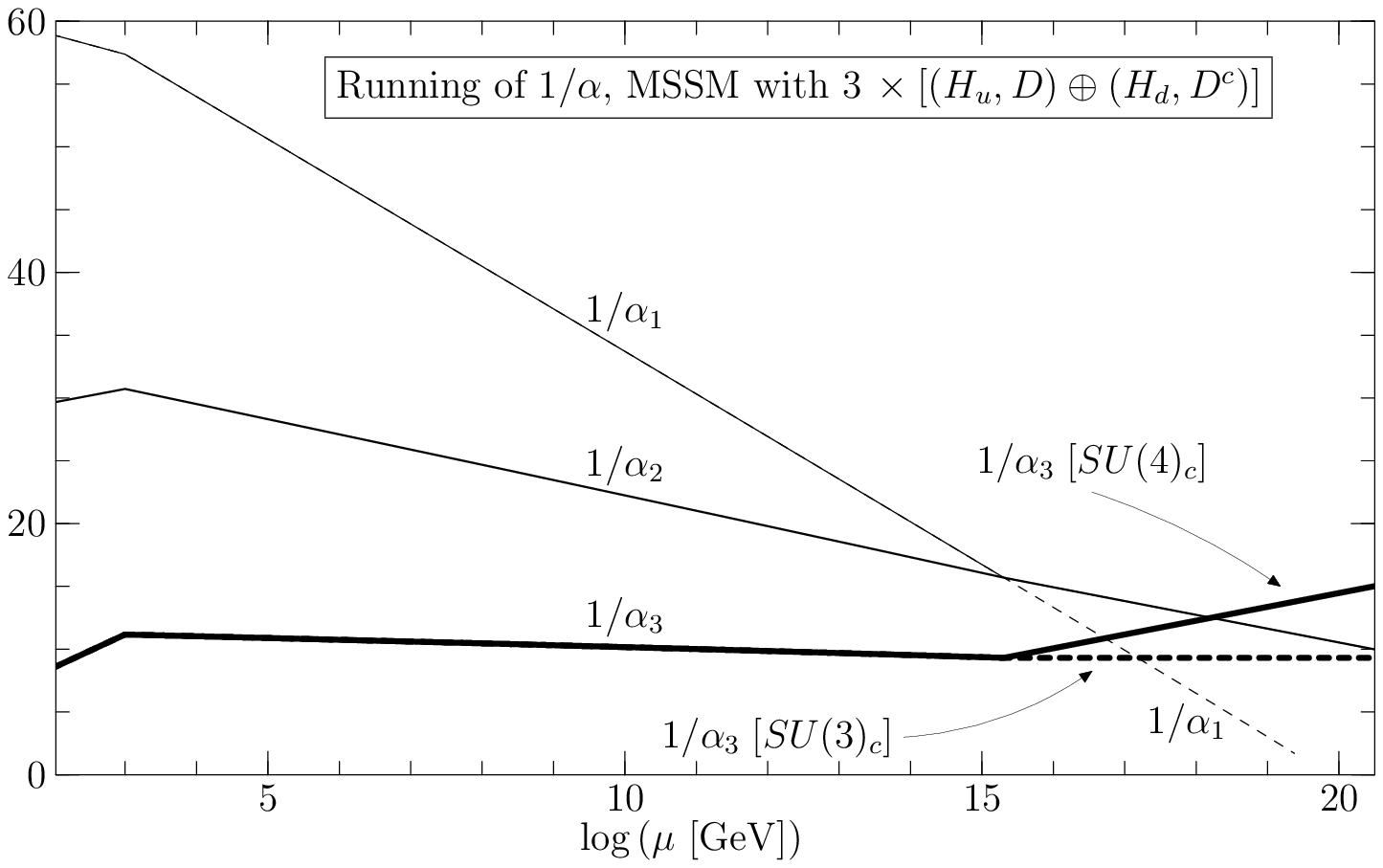}
\end{center}
\vspace{-5mm}
\caption{One-loop running couplings for the MSSM spectrum with one
  (top) or three (bottom) complete matter/Higgs families and
  left-right symmetry above the neutrino mass scale.  Dashed:
  $SU(3)_C\times SU(2)_L\times SU(2)_R$ (SM with left-right symmetry).
  Full: $SU(4)_C\times SU(2)_L\times SU(2)_R$ (Pati-Salam). The
  $SU(2)_L$ is the same in both cases.  
}
\label{fig:uni2}
\end{figure}

Our goal is to construct a model that abandons the doublet-triplet
splitting problem, and solves the $\mu$ problem (i.e. the fact that
the SUSY $\mu$ parameter has to be of order the EW scale for
successful EW symmetry breaking to happen) of the MSSM
simultaneously~\cite{kr_gut}. 
Although our model is motivated more from a bottom-up approach, namely
reconciling the abovementioned model-building perspectives with the
phenomenological demand to observe hints on the GUT structure
directly at LHC, we will describe the model from a top-down view. 


\section{The model from a top-down perspective}  

When keeping the triplet partners of the Higgs doublets in the
low-energy spectrum, there are two important issues: repairing gauge
coupling unification, and much more severe: a mechanism to prevent
too rapid proton decay. Solving the first problem relies on the
observation that in the MSSM spectrum with triplets the $SU(2)$ and
$U(1)$ couplings meet at $10^{14}$ GeV, a natural place for right-handed
neutrinos and the seesaw mechanism. Extending the MSSM symmetry group
to a left-right symmetry $SU(2)_L \times SU(2)_R$ results in a gauge
coupling unification at $10^{21}$ GeV, far above the Planck scale,
cf.~Fig.~\ref{fig:uni2}, dashed lines. But extending $SU(3)_c$ to
$SU(4)_c$ (hence, to a PS group) changes the running of
$\alpha_s$ and allows for a complete unification directly at the
Planck scale (Fig.~\ref{fig:uni2}, full lines). Regarding the second
problem, the standard way to forbid proton decay in the MSSM, $R$
parity, which forbids both diquark and leptoquark couplings for
ordinary quark superfields. For the new triplet Higgses these
couplings are still allowed (either for the scalars or the fermions),
so they must be forbidden by a different mechanism. One way to do is
by imposing flavor symmetries (see below).  

To implement this together with a complete $E_6$ spectrum for the
three generations we start with a particle content of a $N$=2-SUSY
$E_8$ gauge theory. We want to break the $E_8$ down to the $E_6 \times 
SU(3)_F$ combined gauge and flavor symmetry group. Under $N$=2 SUSY
there is a $\rep{248} \oplus \rep{248}$ matter and gauge fundamental
representation of $E_8$. This decomposes as $\rep{248} = \rep{27}_3
\oplus \arep{27}_{\bar{3}} \oplus \rep{78}_1 \oplus \rep{1}_8$ under
$E_6 \times SU(3)_F$. That corresponds to a flavor-triplet of matter
$\rep{27}_3$, a mirror matter multiplet $\arep{27}_{\bar{3}}$, an $E_6$
adjoint $\rep{78}_1$ and the flavor adjoint $\rep{1}_8$
(cf.~e.g.~\cite{Slansky:1981yr}).  

Somewhat below the Planck scale the breaking of $N$=2 SUSY to $N$=1
SUSY removes the mirror matter by an infinite (Kaluza-Klein) tower of
such multiplets and a quartic coupling which in the $E_6$
decomposition contains $(\rep{27}_{3})_i(\arep{27}_{\bar
3})_i(\rep{27}_{3})_j(\arep{27}_{\bar 3})_j$.  An asymmetric spurion
(condensate) $\vev{(\rep{27}_3)_i^a(\arep{27}_{\bar
3})_j^b}=\delta^{ab}\delta_{j,i+1}$ breaks $E_8$ and removes all
mirror matter from the massless spectrum, leaving one zero mode
$(\rep{27}_3)_0$.  The flavor symmetry group is broken by a colorless
spurion, e.g.~$\vev{\rep{1}_8}$. To reduce the symmetry further down
to the PS group, we introduce a spurion
$\vev{\arep{1}_{2,2}\arep{1}_{2,2}}$ which corresponds to a
$\mu$-term-type coupling of mirror-Higgs superfields which occurs in
the decomposition of $\arep{27}_{\bar 3}\arep{27}_{\bar 3}$.  This
also breaks flavor symmetry. An additional allowed spurion would be
$\vev{\arep{27}_{\bar 3}}\sim\vev{\arep{1}_{1,1}}$ which can be used
to distinguish the third generation (by itself, the latter would break
$E_6$ to $SO(10)$, the standard GUT path). 

Similar to the diquark coupling discussed before, the trilinear $E_6$
superpotential $\rep{27}_3\,\rep{27}_3\,\rep{27}_3$ vanishes
identically if flavor symmetry is imposed, so all matter
self-interactions are effectively generated by symmetry breaking.
Looking at other trilinear terms, we can have
$(\rep{27}_3\,\rep{78}_1\,\arep{27}_{\bar 3})$,
$(\rep{78}_1\,\rep{78}_1\,\rep{78}_1)$,
$(\rep{27}_3\,\rep{1}_8\,\arep{27}_{\bar 3})$, and
$\rep{1}_8\,\rep{1}_8\,\rep{1}_{8}$.  The effective superpotential
results from inserting condensates for $\arep{27}_{\bar 3}$ and
integrating out the remaining fields in $\rep{78}_1$ and $\rep{1}_8$.
For these, $E_6\times SU(3)_F$ invariance allows for mass terms.
This construction generates all MSSM superpotential terms, subject to
PS symmetry, as well as couplings $SD^cD$ and $SH_uH_d$.
Concerning baryon number, the only dangerous term is
$\rep{78}_1\,\rep{78}_1\,\rep{78}_1$ which after inserting the
(colorless) condensates into $\rep{27}_3\,\rep{78}_1\,\arep{27}_{\bar
3}$ and integrating out the $\rep{78}_1$ results in additional
trilinear matter couplings.  However, the color-triplet leptoquarks
$X$ contained in the $\rep{78}_1$ do not have a self-coupling: $XXX$
again vanishes by total antisymmetry with respect to all color, left,
and right indices.

\begin{figure}
  \begin{center}
    \includegraphics[width=.25\textwidth]{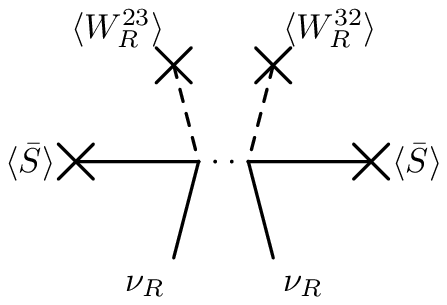}
  \end{center}
  \caption{Two condensates $\vev{\arep{27}_{\bar{3}}}
  \sim \vev{\arep{1}_{1,1}}$ and $\vev{\rep{78}_1}$ generate masses for
  the right-handed neutrinos and break the Pati-Salam symmetry.}
  \label{fig:seesaw}
\end{figure}

At $10^{14}-10^{15}$ GeV, a field with right-handed neutrino quantum
numbers present among the color- and flavorless fields contained in
the $\rep{78}_1 \sim W_R^{23}$ condenses. A quartic term
$(\rep{27}\,\rep{78}\,\arep{27})^2$ in the effective superpotential
generates a right-handed neutrino mass from the diagram in
Fig.~\ref{fig:seesaw}. This is in accordance with the seesaw
mechanism and breaks the PS symmetry down to the SM gauge group.  Near
the electroweak scale, an $S$ condensate generates a $\mu$ term (like
in the NMSSM), and standard radiative breaking of the electroweak
symmetry can occur as in the MSSM.   

At this point we reconcile possible operators mediating proton decay
and how they are forbidden in the model mentioned above. Considering
the exchange of gauge superfields, the proton-decay scale is lifted
together with the GUT scale to $10^{18}$ GeV since PS gauge bosons do
not mediate proton decay.  The PS-invariant superpotential terms
containing $D$ and $D^c$ are
$\arep{4}_{1,2}\rep{6}_{1,1}\arep{4}_{1,2}$ and 
$\rep{4}_{2,1}\rep{6}_{1,1}\rep{4}_{2,1}$.  The first term induces a
leptoquark coupling $u^ce^cD$ together with a diquark coupling
$u^cd^cD^c$, the second term provides leptoquark couplings $D^c\ell q$
and diquark couplings $Dqq$.  Unfortunately, at least one of them is
required to make the $D$ particles decay, so not all of them can be
forbidden (which is on the other hand unnecessary). Here is where the
flavor symmetry comes in. Imposing flavor symmetry on the diquark
couplings on top of $SU(2)_L$ and $SU(2)_R$ symmetry uniquely selects
the structure 
\begin{align*}
  D q_L q_L &= \epsilon^{abc}\epsilon_{\alpha\beta\gamma}\epsilon_{jk}
  D^a_{\alpha} (q_L)^b_{\beta j} (q_L)^c_{\gamma k}
\end{align*}
($D^c$ analogous), where $(a,b,c)$, $(\alpha,\beta,\gamma)$, and
$(i,j,k)$ are indices in flavor, color, and $SU(2)_L$ space,
respectively.  Due to the total antisymmetry of three $\epsilon$
symbols, this term exactly vanishes.  This property continues to hold
if we impose larger gauge symmetries (PS, $SO(10)$, $E_6$) on the
superpotential, as long as $SU(2)_L\times SU(2)_R$ is a subgroup.
Flavor symmetry in conjunction with color and left-right symmetry thus
eliminates diquark couplings, and baryon number automatically emerges
as a symmetry of the superpotential.

While $SU(2)_L$ is exact down to the TeV scale, breaking $SU(2)_R$
and $SU(3)_F$ at high energies might re-introduce $D/D^c$ diquark
couplings and thus proton-decay operators.  To exclude them, we have
to impose baryon number on all $SU(2)_R$- and $SU(3)_F$-breaking
spurions that connect $D$ or $D^c$ with other fields: they have to
involve quark fields in color-singlet pairs (or respect flavor
symmetry).  This is easy to realize since any spontaneous
symmetry-breaking can be associated to condensates at most bilinear in
fields of the fundamental representation.  After integrating out gauge
superfields, baryon number emerges as an exact symmetry of the
low-energy theory.  The flavor symmetry needs not leave obvious
traces.

\section{Phenomenology}

The low-energy spectrum of the abovementioned model is quite similar
to the standard $E_6$ models, containing one to three families of
pairs of Higgses (including their fermionic partners) and the
corresponding leptoquarks at the TeV scale. Furthermore, there are up
to three generations of singlet scalars and their fermionic partners.  
In case of the complete content, there are 6 scalar leptoquarks $D_L$
and $D_R$ and the 3 leptoquarkinos, $\tilde{D}$. In addition to the
MSSM Higgses, 4 charged and 14 neutral (un-)Higgses (for an overview
of non-standard Higgses, see~\cite{review}), 2 additional charginos
and 7 additional neutralinos.  

Concerning the additional singlets, there are two different
possibilities: either there is a gauged $U(1)$ symmetry with a
corresponding $Z'$ gauge boson at the TeV scale which protects the
singlets and also the leptoquarks directly from getting GUT scale
masses. Or this gauged symmetry is broken at the GUT scale but remains
valid as a global symmetry down to the TeV scale. Explicit
symmetry-breaking operators are induced that avoid massless axions. 
At the scale of soft SUSY breaking, VEVs are allowed for the neutral
components of $H_u,H_d$, and for~$S$.  In all non-minimal versions of
the model, these condensates, $\vev{H_u},\vev{H_d},\vev{S}$, are
vectors in family space.  The Higgs and $S$ superfield vectors can be
rotated such that only one component, the third one, gets a VEV and
provides MSSM-like $H_u$ and $H_d$ scalars and higgsinos.  

Yukawa couplings to matter are possible also for the two unhiggs
generations $h_u,h_d,\sigma$ that do not get a
VEV~\cite{Ellis:1985yc}.  To avoid FCNCs via double exchange of
charged unhiggses, the Yukawa matrix entries for them should either be
small or vanish exactly~\cite{Campbell:1986xd}.  The
latter case is equivalent to an extra $Z_2$ symmetry, $H$-parity,
which is odd just for the unhiggs superfields.  Conservation of
$H$-parity would make the lightest unhiggs (or unhiggsino) a
dark-matter candidate~\cite{Griest:1989ew}, adding to the lightest
superparticle (LSP) as the dark-matter candidate of $R$-parity
conservation. Since proton decay is already forbidden by flavor
symmetries, we could alternatively drop lepton number and thus
introduce the full phenomenology of $R$-parity violation, while dark
matter is provided by unhiggses. 

Even if the unhigges $h_u,h_d$ have negligible couplings to ordinary
matter, they can still be pair-pro\-duced at colliders. Their decays
involve ordinary Higgses (including singlets), gauge bosons, or
charginos and neutralinos.  Some of these signals are detectable at
the LHC, all are easily identifiable at the ILC.  Unhiggses could also
occur in decay cascades of higher-level Higgses, charginos, and
neutralinos if kinematically allowed.  Alternatively, if $H$-parity
does not play a role, unhigges may couple significantly to some light
quarks and leptons.  In this case, there is resonant production in
$q\bar q$ annihilation. 

The particles associated with singlet superfields $S$ consist of one
scalar, one pseudoscalar, and one neutralino each.  They are all
neutral and mix with other Higgs and higgsino states.  Production and
decay occurs via mixing only, signals are thus similar to MSSM Higgses
and neutralinos.

The leptoquark superfields $D$ and $D^c$ acquire Dirac masses
proportional to $\vev{S}$.  The masses are considerably enhanced by
renormalization-group running, but some of them could be suppressed by
small Yukawa couplings to~$S$.  Thus, at the LHC we expect up to three
down-type scalar leptoquarks with arbitrary masses.  Depending on the
structure of leptoquark couplings, various decay patterns are
possible.  The most likely variant is dominant coupling to the third
generation, so leptoquarks are pair-produced in $gg$ fusion and decay
into $t\tau$ or $b\nu_\tau$ final states.  They would also show up in
cascades of gluino or squark decay, if kinematically allowed.  The
superpartners (`leptoquarkinos') should be somewhat lighter, decaying
into $\tilde t\tau$, $t\tilde\tau$, $\tilde b\nu_\tau$, or
$b\tilde\nu_\tau$.

The role of flavor symmetry in prohibiting diquark couplings of $D$
fields suggests another scenario: if the dominant terms that induce
leptoquark couplings exhibit flavor symmetry, leptoquark decays
involve all generations simultaneously. This would lead to distinctive
signatures such as $t\mu$, $te$, or light jet plus $\mu$ or $\tau$.
Additional production channels $gq\to D\ell$ would appear.  Analogous
statements hold for the corresponding fermion superpartners~$\tilde D$.

The next step is the calculation of the low-energy spectrum from
certain high-scale boundary conditions
\cite{spectrum}. In~\cite{spectrum} we developed a program for
solving numerically the system of renormalization group equations
(RGEs) of the SUSY and soft breaking parameters, where special care
has to be taken to include the flavor structure of the
equations. E.g.~the RGE for the   Yukawa coupling $Y_{ijk}^{S_H}
\hat{S}_i \hat{H_j^u} \epsilon \hat{H}_k^d$ is $dY_{ijk}^{S_H}/d\log\mu = $
\begin{align*}
  \frac{1}{16\pi^2} \biggl\{ & \;
       \phantom{+} \left( 3 Y^u_{jrs} Y^{u\,\dagger}_{msr} + Y^{S_H}_{rjs}
       Y^{S_H\,\dagger}_{rsm} \right) Y^{S_H}_{imk} \\
       &\;
       + Y^{S_H}_{ijm} \left( 3  Y^{d\,\dagger}_{msr} Y^d_{krs} + Y^{S_H}_{rms}
       Y^{S_H\,\dagger}_{rsk} \right) \\
       &\; + 
       \left( 3 Y^{S_D\,\dagger}_{irs} Y^{S_D}_{msr} + 2
       Y^{S_H\,\dagger}_{irs} Y^{S_H}_{msr} \right) Y^{S_H}_{mjk} 
       \\
       &\; - \left( \tfrac35 g_1^2 + 3 g_2^2 + \tfrac12
       (Q_S^{\prime\,2} + Q_1^{\prime\,2} + Q_2^{\prime\,2}) g_4^2\right) 
       Y^{S_H}_{ijk}
 \biggr\} 
\end{align*}
Here, $Y_u$ and $Y_d$ are the standard MSSM Yukawa couplings for up-
and down-quark superfields, while $Y^{S_D}$ is the coupling for $S_i
D_j D^c_k$. $Q_1'$, $Q_2'$ and $Q_S'$ are the $U(1)'$ charges of the
Higgs doublets and singlet, respectively, while $g_4§$ is the
corresponding gauge coupling. To study the
collider phenomenology of this model, the model is being implemented
(following the conventions of~\cite{SLHA}) into the event generator
WHIZARD~\cite{omega,whizard,omwhiz}, which is well-suited for models
beyond the SM~\cite{omwhiz_resonances,omwhiz_little,omwhiz_bsm}.  

\section{Conclusions}

In conclusion, we have explored a SUSY-GUT scenario without
doublet-triplet splitting, such that the low-energy spectrum contains
color-triplet leptoquarks~$D$ and their superpartners.  The constraint
of gauge-cou\-pling unification then points to the existence of two
distinct high-energy scales.  The first threshold is at
$10^{15}$ GeV where the MSSM gauge group is extended to the
PS group $SU(4)_C\times SU(2)_L\times SU(2)_R$ and
right-handed neutrino masses are generated.  At the higher energy
$10^{18}$ GeV, slightly below the Planck scale, complete unification
(e.g., $E_6$) is located, possibly in the context of a superstring
theory.  

While gauge interactions in a PS GUT do not trigger proton
decay, proton decay via the superpotential can be eliminated by an
underlying flavor symmetry.  $R$-parity conservation is no longer 
mandatory. While the LHC will serve as an ideal machine to discover
especially the colored part of the spectrum of this model, an ILC is
probably needed to disentangle the complex structure of the
weakly-interacting part of that model.

\section{Acknowledgments} 
 
JR was partially supported by the Helmholtz-Gemein\-schaft under
Grant No. VH-NG-005 and the Bundes\-ministerium f\"ur Bildung und
Forschung, Germany, under Grant No. 05HA6VFB.

%
%

\end{document}